\documentclass[aps,twocolumn,superscriptaddress,english,10pt,nofootinbib,preprintnumbers]{revtex4}
\usepackage{amssymb, amsmath, bm, dcolumn, epsf, graphicx, latexsym, slashed, simplewick,mathtools}
\usepackage[utf8]{inputenc}
\usepackage[normalem]{ulem}
\usepackage{color}

\def\be{\begin{equation}}
\def\ee{\end{equation}}
\def\bea{\begin{eqnarray}}
\def\eea{\end{eqnarray}}
\newcommand{\bsm}[1]
\AtBeginDocument{%
  \makeatletter
  \long\def\@makecaption#1#2{%
    \vskip\abovecaptionskip
    {\small\textbf{#1.}~#2\par}%
    \vskip\belowcaptionskip
  }
  \makeatother
}
\usepackage{aas_macros}
\usepackage{amsthm}
\usepackage{float}
\usepackage{comment}
\usepackage[export]{adjustbox}
\usepackage[normalem]{ulem}
\usepackage{hyperref}

\begin{document}

\title{A Cosmological Uncertainty Relation and Late-Universe Acceleration}

\author{Savvas M.~Koushiappas}
\affiliation{Department of Physics and Brown Center for Theoretical
Physics \& Innovation, Brown University, Providence, RI 02912-1843, USA}
\email{koushiappas@brown.edu}

\begin{abstract}

We propose that the size of the universe and its expansion rate cannot be 
simultaneously specified with arbitrary precision—a quantum mechanical uncertainty 
encoded  via a deformed commutation relation for the scale factor. This deformation
introduces a geometric correction to the Friedmann equation, where the resulting 
cosmological dynamics are governed entirely by the sign and magnitude of a single free 
exponent. For a positive exponent, the model predicts late-time dark energy with 
$w>-1$,
leaving a distinct expansion history that is testable by current and 
next-generation large-scale structure surveys. Conversely, a sufficiently negative 
exponent yields a nonsingular classical bounce, resolving the big bang singularity. 
Notably, the model requires no exotic particles or fields and preserves a scale-invariant primordial power spectrum. Rather than operating at the Planck length, this 
deformation naturally manifests as a horizon-scale phenomenon set by the cosmological 
horizon. Within this framework, cosmic acceleration emerges as the macroscopic imprint 
of quantum gravity at the cosmological horizon.

\end{abstract}

\maketitle

\section{Introduction}
\label{sec:intro}

The accelerated expansion of the universe ~\cite{Riess1998,Perlmutter1999}
confirmed by every cosmological probe since its discovery, has no established
microphysical origin. The standard cosmological model, a cosmological constant $\Lambda$ plus cold dark matter ($\Lambda$CDM),  fits the data with a single
free parameter (the scale factor $a$), but the cosmological constant it requires is $122$
orders of magnitude smaller than any natural estimate from quantum
field theory. In addition, recent Baryon Acoustic Oscillation (BAO) measurements from
DESI~\cite{DESI2024} show a mild preference for
dynamical dark energy over a pure~$\Lambda$ cosmological constant.  It seems that the expansion history
we are now measuring is inconsistent with the simplest theory we have.

Quantum gravity is expected to modify cosmology at some scale, and the standard assumption is 
that this scale is the Planck scale. As a result, any corrections are confined to extremely 
early times (as $a\to0$), and invisible today.  That assumption is grounded in dimensional 
analysis, and not in any theorem, and therefore it is worth exploring the question what 
happens if a modification of general relativity does not respect this separation of scales 
between the Planck time and now. After all, this approach is not new in quantum mechanics. 
For example, the Heisenberg uncertainty principle, $[\hat x,\hat p]=i\hbar$, is a statement about what can be
simultaneously known.  Its consequences --- zero-point energy, the
stability of matter --- are visible at every scale, not just at
the quantum mechanical scale where the commutation relationship between $\hat x$ and $\hat p$  was written down. It is therefore natural to ask,  whether such a statement holds for the
universe as a whole.   

In the FLRW minisuperspace~\cite{DeWitt1967,
Misner1969}, the space of all
spatial three-geometries is truncated by imposing homogeneity and isotropy. This leaves 
the scale factor (plus matter) as the only gravitational degree of freedom.  The Einstein-Hilbert action (on an FLRW metric) gives a Lagrangian which is a function of the scale factor (and its derivative), and a conjugate momentum. Canonical quantization 
results in a commutation relationship between the scale factor and its conjugate momentum, namely 
$[\hat{a}, \hat{p}_a] = i \hbar$. Treating the scale factor as a quantized degree of freedom is the standard framework of canonical quantum cosmology (see e.g.,~\cite{Halliwell2009}). 
It is perhaps natural to think of such commutation relationship as something that everything else is 
built on, i.e., treating $i\hbar$ as fixed, and thinking of the scale factor as just another position 
variable. However this is not the case, $i\hbar$ is not derived from any deeper requirement, it is 
just the value obtained from the canonical quantization procedure. Nothing in the structure of the 
theory forces it to be $i\hbar$, and there is no physical reason the geometry of the universe should 
obey the exact same algebra as a particle on a line. As the scale factor is a geometric degree of 
freedom (and not a position coordinate), it is free to behave differently. 

In this paper we pursue this very basic approach, by deforming the velocity--configuration 
commutation relation, $[\hat a,\hat{\dot a}]$.  
The interpretation of this statement is the cosmological analog
of the Heisenberg uncertainty principle: the size of the universe and its rate of expansion 
cannot be simultaneously specified with arbitrary precision. This is not being proposed as a deformation of canonical quantum cosmology. It is instead presented as the quantum realization of a new kinematic principle.

It is natural to question at what scale this relationship is important. 
Quantum gravity is
supposed to be important at the Planck scale, decoupled from anything we can
observe today. But that expectation rests on dimensional analysis,
and the kinematics of the universe as a whole is a different problem
from the low-energy limit of a field theory. The commutation relation
proposed here holds at any value of~$a$, the scale at which it
matters does not need to be Planckian, and as we show a natural candidate for that scale is the cosmological horizon.

The deformation of cosmological commutation relations is not a new
idea, but the specific suggestion made here --- deforming the
velocity--configuration bracket $[\hat a,\hat{\dot a}]$ rather than
the canonical bracket --- is, to our knowledge, new.  In
Generalized Uncertainty Principle (GUP)
models~\cite{Maggiore1993,Das2008} the deformation acts on
$[\hat a,\hat p_a]$ and places the correction on the right-hand
side of the Friedmann equation as a function of the energy
density~\cite{Vakili2008,Battisti2008,Bosso2020}.
Garc{\'\i}a-Compe{\'a}n, Obreg{\'o}n, and
Ram{\'\i}rez~\cite{GarciaCompean2002} pioneered the replacement of
the pointwise product by the Moyal star product in the
Wheeler--DeWitt equation, implementing $[\hat a,\hat\phi]=i\theta$
between the scale factor and an auxiliary scalar field; subsequent
work extended this to multi-field minisuperspaces and anisotropic
models~\cite{Guzman2007,Obregon2001,Vakili2012,Barbosa2004,Bastos2008}.
In loop quantum cosmology (LQC) the holonomy corrections yield
$H^{2}=(8\pi G/3)\rho(1-\rho/\rho_c)$~\cite{Ashtekar2006,Singh2009},
with the bounce arising from a $\rho^{2}/\rho_c$ term on the
right-hand side~\cite{Ashtekar2011,Bojowald2001}; perturbation
theory has been developed
extensively~\cite{Cailleteau2012,FernandezMendez2012,Barrau2015}.
Phase-space noncommutativity, $[\hat p_a,\hat p_\phi]=i\eta$, has
also been studied~\cite{Obregon2001,Vakili2010,DiazBarron2021,DiazBarron2019},
and noncommutative cosmology has been extended to Brans--Dicke
theory~\cite{Rasouli2011,Rasouli2014}, $\kappa$-deformed
spacetime~\cite{Gubitosi2023}, and branch-cut quantum
gravity~\cite{Zen2023}.

The present proposal is structurally distinct from all of these.
Because $\hat{\dot a}$ is not the canonical momentum but the
velocity, deforming $[\hat a,\hat{\dot a}]$ is {\it not} equivalent 
to a GUP-type deformation of $[\hat{a}, \hat{p}_a]$; the induced canonical 
bracket has a configuration-dependent right-hand side rather than a 
momentum-dependent one, and no canonical transformation interchanges the two. 
As a result, the correction appears on the left-hand side as a
geometric modification to the expansion rate itself, and not on the
right as an effective energy density.

Most modified-gravity studies begin by writing down a Lagrangian, 
deriving equations of motion, and analyzing the resulting cosmology. 
The present approach deforms the kinematic structure—the operator algebra of the gravitational degrees of freedom—rather than the dynamics. Once the deformed algebra is written down, the modified 
Friedmann equation is not chosen but follows from operator self-adjointness, 
the operator  uncertainty principle, and the FLRW Hamiltonian constraint. 
A Hamiltonian formulation on the deformed phase space then re-emerges 
as a consistency check  rather than as an input -- the 
resulting geometric correction is forced by the algebra.  This is in 
keeping with the Heisenberg analogy that motivated the deformation: the 
uncertainty principle is not a statement about a Lagrangian, but about 
what can be simultaneously known.

The paper is organized as follows.  Section~\ref{sec:algebra_rep}
establishes the operator representation of the algebra, resolves
the ordering ambiguity by self-adjointness, and derives the
deformed momentum operator.  Section~\ref{sec:friedmann} derives
the modified Friedmann equation as a consequence of the deformed
uncertainty relation and verifies its Hamiltonian formulation.
 Section~\ref{sec:WdW} derives the
modified Wheeler--DeWitt equation.  Sections~\ref{sec:earlyuniv}
and~\ref{sec:late} work out the early- and late-universe
consequences, including the cosmological regime
$a_0\sim H_0^{-1}$, $\beta\ll H_0$ and its observational
predictions.  Section~\ref{sec:related} positions the proposal
within the existing literature.  Section~\ref{sec:discussion}
discusses the model's limitations, open questions, and physical interpretation. We 
conclude in Section~\ref{sec:conclusions}.  
Throughout we set $k=0$ (spatially flat universe) and
use the metric signature $(-,+,+,+)$.

\section{The Cosmological Uncertainty Relation and its Operator Representation}
\label{sec:algebra_rep}

We propose the following deformation of the velocity--configuration
algebra.  The commutator $[\hat a,\hat{\dot a}]$ has dimensions
$L^{2}T^{-1}$.  With $[\beta]=T^{-1}$ and $[a_0]=L$, the combination
$\beta\hat a^{2}(1+(\hat a/a_0)^{n})$ has the same dimensions
for any real~$n$, since $(a/a_0)^{n}$ is dimensionless.  The noncommuctative (NC) algebra 
\begin{equation}
[\hat a,\hat{\dot a}]
  = -i\beta\,\hat a^{2}\!\left[1+\Bigl(\frac{\hat a}{a_0}\Bigr)^{\!n}
    \right],
\label{eq:algebra}
\end{equation}
is dimensionally consistent for any real~$n$.  The normalization is 
provided by~$\hat a^{2}$, the minimal
polynomial in~$\hat a$ with the correct dimensions. 
The
parameter~$a_0$ sets the crossover scale between the constant and
power-law regimes of the deformation.

The scale factor is positive-definite, so we work in the configuration-space representation 
$\hat a\,\Psi(a)=a\,\Psi(a)$, with $a>0$
and  normalizable wave functions.  Writing
$F(a)\equiv1+(a/a_0)^{n}$, the one-parameter family of operators
\begin{equation}
\hat{\dot a}^{(s)}
  = i\beta\,\hat a^{2-s}\frac{d}{da}\hat a^{s}\cdot F(\hat a)\,,
  \qquad s\in[0,2]\,,
\label{eq:adots}
\end{equation}
all satisfy~(\ref{eq:algebra}) independently of~$s$, since the
commutator depends only on the total number of powers of~$\hat a$
on either side of the derivative: $[\hat a,\hat{\dot a}^{(s)}]=-i\beta\hat
a^{2}F$, for every~$s$.

The ordering ambiguity is resolved by requiring $\hat{\dot a}$ to
be self-adjoint.  The physical Hubble rate involves $\dot{a}^{2}/a^{2}$,
which must have a non-negative expectation value, so this
requirement is physical rather than a convention.  Computing the
adjoint of $\hat{\dot a}^{(s)}$ for the case $F = 1$ gives
\begin{equation}
\bigl(\hat{\dot a}^{(s)}\bigr)^{\dagger}
  = \hat{\dot a}^{(2-s)}\,,
\end{equation}
so that self-adjointness holds if and only if $s = 1$.  This same
constraint applies for general $F(a) > 0$: when $F$ is split
symmetrically as $\sqrt{F}\,(\cdot)\,\sqrt{F}$ around the
differential, the integration-by-parts argument leading to
the conjugation $s \to 2-s$ is unchanged, and $s = 1$ remains the
unique self-adjoint ordering.  Asymmetric placements of $F$ do
not preserve the structure of the operator under conjugation
and are excluded.  The resulting operator, unique up to additive functions of $\hat a$ that vanish in the classical limit,  is
\begin{equation}
\hat{\dot a}
  = i\beta\,\sqrt{F(\hat a)}\,\hat a\,\frac{d}{da}\,
    \hat a\,\sqrt{F(\hat a)}\,.
\label{eq:adot}
\end{equation}
Applying Eq.~\ref{eq:adot} to a wavefunction and expanding the derivatives,
\begin{equation}
\hat{\dot a}\,\Psi
  = i\beta\!\left(aF\,\Psi
    +\tfrac{1}{2}a^{2}F'\,\Psi
    +a^{2}F\,\Psi'\right),
\label{eq:adotexpand}
\end{equation}
where $F'\equiv dF/da$.  One verifies that
$[\hat a,\hat{\dot a}]=-i\beta a^{2}F$ still holds. 

The self-adjointness of $\hat{\dot a}$ follows from a single
integration by parts.  Using Eq.~(\ref{eq:adot}),
\begin{equation}
\langle\varphi|\hat{\dot a}\,\psi\rangle
  = i\beta\!\int_{0}^{\infty}\!\bar\varphi\,\sqrt{F}\,a\,
    \frac{d}{da}\!\left[a\,\sqrt{F}\,\psi\right]\,da\,.
\end{equation}
Integrating by parts moves the derivative onto $\bar\varphi$:
\begin{equation}
\langle\varphi|\hat{\dot a}\,\psi\rangle
  = \left[a^{2}F\,\bar\varphi\,\psi\right]_{0}^{\infty}
    - i\beta\!\int_{0}^{\infty}\!a\,\sqrt{F}\,\psi\,
    \frac{d}{da}\!\left[\bar\varphi\,\sqrt{F}\,a\right]\,da\,.
    \label{eq:ibp}
\end{equation}
The boundary term vanishes at $a = 0$ because of the $a^{2}$ factor
and at $a \to \infty$ by normalizability\footnote{Throughout this paper, normalizability is 
meant in the standard sense for cosmological wave functions: square-integrable with sufficient 
pointwise decay at infinity.} 
of $\bar\varphi\,\psi$.
The remaining integral is precisely the conjugate of
$\langle\hat{\dot a}\,\varphi|\psi\rangle$, since $\sqrt{F}\,a$ is
real and conjugation flips the sign of $i\beta$.  Therefore
\begin{equation}
(\hat{\dot a})^{\dagger} = \hat{\dot a}
  \qquad\text{for all } F(a) > 0\text{ and } a > 0\,.
\label{eq:adotsa}
\end{equation}

The operator~(\ref{eq:adot}) is a kinematic object defined
independently of the Hamiltonian; it is not the Heisenberg-picture
time derivative
$\widehat{da/dt}\equiv-i[\hat a,\hat{\mathcal{H}}_{\mathrm{def}}]$,
which requires a Hamiltonian to be specified.  We interpret
$\hat{\dot a}$ as a \emph{deformation generator}: an operator
encoding the NC structure of the minisuperspace phase space that
coincides with $\widehat{da/dt}$ only in the undeformed case.

To connect with the Hamiltonian formulation of cosmology we switch variables from $\dot{a}$ to 
the canonical momentum $p_a$. We do this in the following way:
The classical limit replaces operators by their values, with the
commutator mapping to the Poisson bracket via
$[\,\cdot\,,\cdot\,]\to i\{\,\cdot\,,\cdot\,\}$ (with $\hbar=1$).
Equation~\eqref{eq:algebra}  gives the deformed classical
Poisson bracket
\begin{equation}
\{a,\dot a\}_{\mathrm{def}}
  = -\beta\,a^{2}\!\left[1+\Bigl(\frac{a}{a_0}\Bigr)^{\!n}\right].
\label{eq:poisson}
\end{equation}
The reduced FLRW Lagrangian
${\cal L} = -(3/8\pi G)\,a\dot a^{2} + a^{3}{\cal L}_{\mathrm{matter}}$
gives $p_a = -(3/4\pi G)\,a\dot a$, and the Leibniz rule applied
to~\eqref{eq:poisson} yields the deformed bracket 
\begin{equation}
\{a,p_a\}_{\mathrm{def}}
  = \frac{3\beta}{4\pi G}\,a^{3}\!\left[1+
    \Bigl(\frac{a}{a_0}\Bigr)^{\!n}\right].
\label{eq:defpb}
\end{equation}
Promoting to operators via $\{\,\cdot\,,\cdot\,\}\to-i[\,\cdot\,,\cdot\,]$,
\begin{equation}
[\hat a,\hat p_a^{(\mathrm{def})}]
  = \frac{3i\beta}{4\pi G}\,\hat a^{3}\!
    \left[1+\Bigl(\frac{\hat a}{a_0}\Bigr)^{\!n}\right]
  \equiv i\,\omega(\hat a)\,.
\label{eq:defcomm}
\end{equation}
where
\begin{equation}
\omega(a)\equiv Ka^{3}F(a)\,,
\qquad K\equiv\frac{3\beta}{4\pi G}\,.
\label{eq:omegadef}
\end{equation}
Equation~\eqref{eq:defcomm} replaces $[\hat a,\hat p_a]=i$ throughout the deformed theory.
The deformed commutator is a pure function of~$a$ with no state
dependence.

We will now calculate the explicit form of $\hat p_a^{(\mathrm{def})}$
satisfying~\eqref{eq:defcomm} that is also self-adjoint.  The
analysis follows the same path as the construction of
$\hat{\dot a}$.
We start by choosing the candidate momentum operator 
\begin{equation}
\hat p^{(\mathrm{c})}
  \equiv -i\,\omega(\hat a)\,\partial_{a}\,.
\end{equation}
This satisfies the deformed commutator by direct computation:
\begin{equation}
[\hat a,\,\hat p^{(\mathrm{c})}]\,\Psi
  = a\bigl(-i\omega\Psi'\bigr)
    - \bigl(-i\omega\bigr)(a\Psi)'
  = i\,\omega(a)\,\Psi\,.
\end{equation}
However, $\hat p^{(\mathrm{c})}$ is not self-adjoint.
Integration by parts on $\langle\varphi|\hat p^{\mathrm{c}}\psi\rangle$
moves the derivative onto $\bar\varphi$ (as in Eq.~(\ref{eq:ibp})) but here it  generates an extra term from the derivative acting on $\omega$:
\begin{equation}
(\hat p^{(\mathrm{c})})^{\dagger}
  = -i\,\omega(\hat a)\,\partial_{a} - i\,\omega'(\hat a)\,,
\end{equation}
where $\omega' \equiv d\omega/da$.
The boundary term vanishes for the same reason as in the
$\hat{\dot a}$ analysis: $\omega(a) = K a^{3} F$ contains a
factor~$a^{3}$ that forces it to zero at $a = 0$, and
normalizability is responsible for driving the integral to zero at infinity as before.

Self-adjointness is restored by symmetrizing.  Just as the
symmetric $\sqrt{F}\,(\cdot)\,\sqrt{F}$ placement made
$\hat{\dot a}$ self-adjoint by  splitting $F$
across the differential, the symmetric combination of
$\hat p^{(\mathrm{c})}$ with its adjoint produces a
self-adjoint operator:
\begin{equation}
\hat p_a^{(\mathrm{def})}
  = \frac{1}{2}\bigl[\hat p^{(\mathrm{c})}
    + (\hat p^{(\mathrm{c})})^{\dagger}\bigr]
  = -i\!\left[\omega(\hat a)\,\partial_{a}
    + \frac{1}{2}\,\omega'(\hat a)\right].
\label{eq:pdef}
\end{equation}
The coefficient of~$1/2$ in the last term is uniquely fixed by self-adjointness and any other value would make the operator non-Hermitian\footnote{ Consider $\hat p_\alpha = -i\omega\,\partial_a - i\alpha\,\omega'$. It  has adjoint $-i\omega\,\partial_a + i(\alpha-1)\,\omega'$, so
$\hat p_\alpha^\dagger = \hat p_\alpha$ if only $\alpha = 1/2$.}.
Equation~\eqref{eq:pdef} still satisfies~\eqref{eq:defcomm},
since $-\tfrac{i}{2}\,\omega'$ is a function of $\hat a$ alone
and commutes with~$\hat a$, leaving the commutator unchanged.

The deformed momentum operator~\eqref{eq:pdef} has a  canonical-quantization 
interpretation. The symmetric (Weyl) ordering of the classical relation $p_a = -(3/4\pi G)\,a\dot{a}$  gives
\begin{equation}
\hat{p}_a^{(\rm def)} = -\frac{3}{4\pi G}\cdot\frac{1}{2}(\hat{a}\hat{\dot{a}} + \hat{\dot{a}}\hat{a}), 
\end{equation}
which is the same as equation~\eqref{eq:pdef} as one can verify by direct computation.
The bracket-based construction and the Weyl-ordered classical product are independent 
prescriptions, and their agreement confirms that the operator content of 
$\hat{p}_a^{(\rm def)}$ is uniquely determined and no operator-ordering freedom remains.

\section{The Modified Friedmann Equation}
\label{sec:friedmann}

The deformed commutation relation~Eq.~(1) implies a modified uncertainty relation
between~$a$ and~$p_a$.  In this section we show that the
irreducible quantum variance forced by this relation adds a
geometric correction to the Friedmann equation.

As was discussed earlier, in the standard FLRW Hamiltonian formulation of cosmology,  the canonical momentum is
$p_a=-(3/4\pi G)\,a\dot a$, so the Hubble rate is
$H=\dot a/a=-(4\pi G/3)\,p_a/a^{2}$.
The Friedmann equation is then the Hamiltonian constraint
$\mathcal{H}=0$, which in terms of $a$ and $p_a$ is 
\begin{equation}
\frac{16\pi^{2}G^{2}}{9a^{4}}\,p_a^{2}=f(a)\,,
\label{eq:classconstraint}
\end{equation}
with $f(a)\equiv(8\pi G/3)\rho+\Lambda c^{2}/3$. In the classical
theory the Friedmann equation is then given by $H^{2}=f(a)$.

In the deformed theory $[\hat a, \hat p_{a}^{(\mathrm{def})}]
= i\omega(\hat a)$. We will connect this commutation relation to a quantum correction to the Hubble rate. 
We begin with the uncertainty relation 
\begin{equation}
\Delta a\,\Delta p_{a} \geq \frac{1}{2}
  \bigl|\langle[\hat a, \hat p_{a}^{(\mathrm{def})}]\rangle\bigr|
  = \frac{1}{2}\langle\omega(\hat a)\rangle
  \approx \frac{\omega(a)}{2}\,,
\label{eq:robertson}
\end{equation}
where the last step uses
$\langle\omega(\hat a)\rangle \approx \omega(\langle\hat a\rangle)$
for states whose quantum spread is small compared to the mean\footnote{The Taylor expansion
$\langle\omega(\hat a)\rangle = \omega(\langle\hat a\rangle)
+ \tfrac{1}{2}\omega''(\langle\hat a\rangle)(\Delta a)^{2}
+ \mathcal{O}(\Delta a^{3})$ shows that the replacement is
valid up to relative corrections of order
$(\Delta a/\langle\hat a\rangle)^{2}$ (the standard
semiclassical regime).}.
Squaring and solving for $\Delta p_{a}$:
\begin{equation}
(\Delta p_{a})^{2} \geq \frac{\omega(a)^{2}}{4(\Delta a)^{2}}\,.
\label{eq:dpsq}
\end{equation}

The full quantum expectation value of $\hat p_{a}^{2}$
is  $\langle\hat p_{a}^{2}\rangle
= \langle\hat p_{a}\rangle^{2} + (\Delta p_{a})^{2}$, so 
substituting into the LHS of the Hamiltonian
constraint~\eqref{eq:classconstraint} gives a classical part plus an
irreducible quantum correction,
\begin{equation}
\frac{16\pi^{2}G^{2}}{9 a^{4}}\langle\hat p_{a}^{2}\rangle
  = H^{2} + \delta H^{2}\,,
\quad
\delta H^{2} \equiv \frac{16\pi^{2}G^{2}}{9 a^{4}}(\Delta p_{a})^{2}\,.
\label{eq:dHdef}
\end{equation}
Here $H = -(4\pi G/3)\langle\hat p_{a}\rangle/a^{2}$
is the classical Hubble rate and $\delta H^{2}$ is the
irreducible quantum variance introduced by the commutation relation~\eqref{eq:algebra}.

Combining~\eqref{eq:dpsq} and~\eqref{eq:dHdef}, and
substituting $\omega = (3\beta/4\pi G)\,a^{3}F$, we get 
\begin{equation}
\delta H^{2}
  \geq  \frac{\beta^{2}F(a)^{2}}{4\bigl(\Delta a/a\bigr)^{2}}\,,
\label{eq:deltaHbound}
\end{equation}
Note that the quantum correction to~$H^{2}$ is set by the deformation
parameter~$\beta$, the NC profile~$F(a)$, and the relative
spread~$\Delta a/a$.  
As with the zero-point energy of a harmonic oscillator, $\Delta a/a = 1/2$
 gives the smallest correction, $\delta H^2 = \beta^2 F(a)^2$.

The constraint $\mathcal{H}=0$ at the semiclassical level is then
$H^{2}+\beta^{2}F^{2}=f(a)$.  Identifying
$H$ with the observed Hubble rate $H=\dot a/a$
gives the modified Friedmann equation:
\begin{equation}
H^{2}+\beta^{2}\!\left[1+\Bigl(\frac{a}{a_0}\Bigr)^{\!n}\right]^{2}
  =\frac{8\pi G}{3}\,\rho+\frac{\Lambda c^{2}}{3}\,.
\label{eq:modfried}
\end{equation}
This is the main result of this paper. The NC correction $\beta^{2}F^{2}$ is on the left-hand side, 
as a geometric contribution that reduces the
expansion rate available from a given energy budget (set by the RHS of~\refeq{eq:modfried}).  It represents
the minimum quantum variance of the Hubble rate: the deformed
commutation relation prevents $\langle\hat p_a^{2}\rangle$ from being entirely
classical, and the missing expansion rate is the kinematic price
of noncommutativity.  This is analogous to the zero-point energy
of a quantum oscillator, forced by $[\hat x,\hat p]=i\hbar$ and
irremovable by any choice of state.  The correction appears on the
left because it is a property of the kinematics --- the deformed
commutator --- not of the energy content.  Whereas a function of $\rho$ added to the right-
hand side modifies the gravitational response to a given matter content, the correction 
obtained here modifies the kinematics of the scale factor itself; the magnitude of the 
correction depends on $a$, not on the matter content of the universe.

Differentiating~(\ref{eq:modfried}) with respect to~$t$ and using
the continuity equation $\dot\rho+3H(\rho+p/c^{2})=0$ gives the
modified Raychaudhuri equation:
\begin{eqnarray}
\frac{\ddot a}{a}
  = &-&\frac{4\pi G}{3}\!\left(\rho+\frac{3p}{c^{2}}\right)
    +\frac{\Lambda c^{2}}{3} \\
    &-&\beta^{2}F(a)\!\left[1+(n+1)\Bigl(\frac{a}{a_0}\Bigr)^{\!n}
    \right],
\label{eq:raychaudhuri}
\end{eqnarray}
where the bracket equals $F+aF'=1+(n+1)(a/a_0)^{n}$, with $F'\equiv dF/da$.  For $n>-1$ the NC term is
always decelerating.  For $n<-1$ it changes sign at
$a_{*}=a_0|n+1|^{1/|n|}$: decelerating at $a<a_{*}$ and
accelerating at $a>a_{*}$ without a cosmological constant.

As a consistency check, we verify that~(\ref{eq:modfried}) admits
a standard Hamiltonian formulation on the deformed phase space.
The commutation relation of \eqref{eq:algebra} implies a deformed bracket
$\{a,p_a\}_{\mathrm{def}}=\omega(a)$ with $\omega=Ka^{3}F$.
Hamilton's equations with lapse~$N$ are
$\dot a=N\omega\,\partial\mathcal{H}/\partial p_a$ and
$\dot p_a=-N\omega\,\partial\mathcal{H}/\partial a$.
The Hamiltonian constraint\footnote{Unlike the undeformed case, $N$ is 
not an independent Lagrange multiplier on the deformed phase space: it is fixed by the
kinematic condition $\dot a = aH$ to the value $N = -1/[2\beta F(a)]$.  The 
deformed constraint $\mathcal{H}_{\mathrm{def}} = 0$ is therefore postulated than 
obtained by varying with respect to a free lapse.} given by
\begin{equation}
\mathcal{H}_{\mathrm{def}}(a,p_a)
  =\frac{16\pi^{2}G^{2}}{9a^{4}}\,p_a^{2}
   +\beta^{2}F(a)^{2}-f(a)=0
\label{eq:hamdef}
\end{equation}
reproduces~(\ref{eq:modfried}) on the constraint surface.
Requiring $\dot a=aH=-(4\pi G/3)\,p_a/a$ determines the lapse,
$N=-1/[2\beta F(a)]$,
which is finite and non-degenerate for all $a>0$.

Setting $\mathcal{H}_{\mathrm{def}}=0$ gives
$p_a^{2}=C(a)$ with
\begin{equation}
C(a)=\frac{9a^{4}}{16\pi^{2}G^{2}}
  \bigl[f(a)-\beta^{2}F(a)^{2}\bigr].
\label{eq:Cdef}
\end{equation}
$C(a)\ge0$ is equivalent to $H^{2}\ge0$; its zeros are the turning
points of the classical dynamics, and their structure --- isolated
points, connected intervals, or one-sided boundaries ---
determines whether the model admits bounces, re-collapse, or
tunneling.

The squared NC term decomposes as
\begin{equation}
\beta^{2}F(a)^{2}
  = \beta^{2}\!\left[1
    + 2\!\left(\frac{a}{a_{0}}\right)^{\!n}
    + \!\left(\frac{a}{a_{0}}\right)^{\!2n}\right],
\label{eq:F2expand}
\end{equation}
into a constant piece $\beta^{2}$, which shifts the effective
cosmological constant to
\begin{equation}
\Lambda_{\mathrm{eff}} = \Lambda - \frac{3\beta^{2}}{c^{2}},
\label{eq:Lambdaeff}
\end{equation}
and two power-law terms which act as effective fluids with equations
of state
\begin{equation}
w_{1} = -1 - \frac{n}{3}\,, \qquad
w_{2} = -1 - \frac{2n}{3}\,.
\label{eq:eos}
\end{equation}
Both are phantom ($w < -1$) for $n > 0$ and non-phantom for $n < 0$.
The shift to $\Lambda_{\mathrm{eff}}$ is large for Planck-scale
$\beta$, inheriting the cosmological constant problem; the model's
options for handling this fine-tuning are discussed in
Section~\ref{sec:discussion_limits}. The crossover scale $a_{0}$ (a
free parameter) marks the transition where the constant and
power-law contributions to $F^{2}$ are equal; for $n > 0$ and $a_{0}
\sim H_{0}^{-1}$ this crossover occurs at the present epoch,
providing a natural explanation for the current onset of acceleration
(see Sec.~\ref{sec:late}).

\section{THE MODIFIED WHEELER--DEWITT EQUATION}
\label{sec:WdW}

The classical constraint $p_a^{2} = C(a)$ of
Eq.~\eqref{eq:Cdef} becomes, upon canonical quantization, a
second-order differential equation for the wave function of the
universe.  In this section we derive its explicit form in the
deformed theory and verify that the modified Friedmann equation is
recovered at leading order in the Wentzel–Kramers–Brillouin (WKB) approximation.  Throughout we work in Planck units
($\hbar = c = 1$), measuring the scale factor in units of
$\ell_P = \sqrt{G}$.

The self-adjoint momentum operator~\eqref{eq:pdef} can be written as 
\begin{equation}
\hat p_a^{(\mathrm{def})}
  = -i\,\omega^{1/2}\,\partial_a\,\omega^{1/2}\,,
\label{eq:pfactor}
\end{equation}
which one verifies by direct expansion: the derivative acting on
$\omega^{1/2}\Psi$ produces a term $\omega'\Psi/(2\omega^{1/2})$,
and multiplying by the outer $\omega^{1/2}$ reproduces
$-i(\omega\Psi' + \tfrac{1}{2}\omega'\Psi)$.  Squaring~\eqref{eq:pfactor},
\begin{equation}
\hat p^{2}\Psi
  = -\,\omega^{1/2}\,\partial_a
    \bigl[\omega\,\partial_a(\omega^{1/2}\Psi)\bigr].
\label{eq:psq}
\end{equation}
Expanding the nested derivatives generates four terms:
$\omega^{2}\Psi''$, two copies of $\omega\omega'\Psi'$, and two
$\Psi$-terms proportional to $\omega\omega''$ and $(\omega')^{2}$.
Using the identity
$\partial_a(\omega^{2}\Psi') = 2\omega\omega'\Psi' + \omega^{2}\Psi''$
to collect the derivative-carrying pieces gives
\begin{equation}
\hat p^{2}\Psi
  = -\partial_a(\omega^{2}\Psi')
    - U_{\mathrm{ord}}(a)\,\Psi\,,
\label{eq:psqfinal}
\end{equation}
with the ordering potential
\begin{equation}
U_{\mathrm{ord}}(a)
  \equiv \frac{1}{2}\,\omega\,\omega''
    + \frac{1}{4}(\omega')^{2}\,.
\label{eq:Vord}
\end{equation}
The quantum Hamiltonian constraint
$\hat p^{2}\Psi = C(a)\Psi$ then is
\begin{equation}
\partial_a\bigl(\omega^{2}\partial_a\Psi\bigr)
  + U_{\mathrm{ord}}(a)\,\Psi + C(a)\,\Psi = 0\,,
\label{eq:WdW}
\end{equation}
which is the modified Wheeler--DeWitt equation.  The kinetic term
$\partial_a(\omega^{2}\partial_a)$ carries the full $a$-dependence
of the deformation function; the ordering potential
$U_{\mathrm{ord}}$ encodes the operator-ordering choice made
in~\eqref{eq:pfactor}.

The explicit form of~\eqref{eq:WdW} follows from computing
$\omega'$ and $\omega''$ for $\omega = Ka^{3}F$ with
$K = 3\beta/(4\pi G)$ and $F(a) = 1 + (a/a_0)^{n}$.  Direct
differentiation gives
$\omega' = Ka^{2}(3F + aF')$ and
$\omega'' = Ka(6F + 6aF' + a^{2}F'')$.  Substituting
into~\eqref{eq:Vord}, dividing by $K^{2}$, and using
$C(a)/K^{2} = a^{4}[f(a) - \beta^{2}F^{2}]/\beta^{2}$, the Wheeler-DeWitt 
equation~\eqref{eq:WdW} takes the form
\begin{widetext}
\begin{equation}
\partial_a\bigl(a^{6}F^{2}\,\partial_a\Psi\bigr)
  + a^{4}\!\left[\frac{21}{4}F^{2} + \frac{9}{2}aFF'
    + \frac{a^{2}}{2}FF'' + \frac{a^{2}}{4}(F')^{2}\right]\Psi 
  = -\,\frac{a^{4}\bigl[f(a) - \beta^{2}F^{2}\bigr]}{\beta^{2}}\,\Psi\,.
\label{eq:WdWexplicit}
\end{equation}
\end{widetext}
The kinetic prefactor $a^{6}F^{2}$ is set entirely by the
deformation function $\omega(a)$. 
In the limit $F \to 1$ and $F', F'' \to 0$, the kinetic operator reduces to  
$\partial_a(a^6\,\partial_a)$~\cite{DeWitt1967,Misner1969,Halliwell2009}, together with an 
additive ordering potential $(21/4)a^4$ that follows from the symmetric construction 
of~\eqref{eq:pdef} in Section~\ref{sec:algebra_rep},  while the 
right-hand side reduces correspondingly to $-a^4 f(a)/\beta^2$ (with $\beta^2$
setting the units of the canonical bracket).

The semiclassical limit of~\eqref{eq:WdWexplicit} recovers the
modified Friedmann equation.  To show this, substituting the WKB ansatz
$\Psi(a) = A(a)\,e^{iS(a)}$ where $S(a)$ is the classical action (with $S'(a) = p_a/\omega(a)$ 
at leading order) and $A(a)$ is the amplitude. Then,  the real part at leading order in the gradient expansion
gives $\omega^{2}(S')^{2} = C(a)$.  Since
$\hat p\Psi \approx \omega\,S'\,\Psi$ at this order, this is
$p_a^{2} = C(a)$, equivalent to
$H^{2} = f(a) - \beta^{2}F(a)^{2}$, which is the same as  the modified Friedmann
equation.  The ordering potential $U_{\mathrm{ord}}(a)$ contributes only at
subleading order in the WKB expansion and doe snot affect the classical Friedmann equation. 

Note that the limit $\beta\to0$ should not be
interpreted as a return to canonical quantum cosmology. Unlike GUP-type
constructions, where a deformation is added to a pre-existing quantum algebra, 
Eq.~\eqref{eq:algebra} is the fundamental kinematic statement of the theory. As 
$\beta\to0$, the cosmological uncertainty relation disappears, the minisuperspace 
algebra becomes commutative, and the theory approaches classical FLRW dynamics rather 
than the conventional Wheeler-DeWitt framework.

\section{EARLY-UNIVERSE CONSEQUENCES}
\label{sec:earlyuniv}

The model has two observationally distinct parameter regimes. In this section we discuss the 
\emph{Planck regime} ($a_0 = \ell_P$, $\beta = t_P^{-1}$); the cosmological regime is treated 
in Section~\ref{sec:late}. 

In the Planck regime, the behavior of the commutation relation in 
the early universe is set by
the sign and magnitude of~$n$.  There are three cases that have interesting implications: 
\begin{itemize} 
\item For $n < -2$ the NC correction grows at small
scale factors, providing a restoring force that produces a
genuine classical bounce (thus eliminating the Big Bang singularity).
\item For $n > 0$ the NC correction
grows with~$a$ and produces a maximum scale factor in the
radiation-dominated epoch, beyond which expansion is classically
forbidden; the Big Bang singularity persists classically, with
quantum tunneling as the only mechanism for crossing the
forbidden region.
\item For $-2 \le n \le 0$ the NC
correction either dilutes or remains bounded at small~$a$, and
the singularity is not resolved. 
\end{itemize}
We analyse the first two dynamically interesting cases below.

A classical bounce requires $H^{2}(a)$ to vanish at some
$a_{\mathrm{bounce}} > 0$ separating a classically forbidden
region $H^{2} < 0$ at smaller scale factors from an allowed
region $H^{2} > 0$ at larger scale factors.  Generic bounces
appear as simple zero crossings of $H^{2}(a)$ with positive
slope, $(d/da)H^{2}|_{a_{\mathrm{bounce}}} > 0$, and $\ddot a > 0$
at the crossing guarantees re-expansion.  The marginal
configuration in which $H^{2}$ touches zero tangentially ---
with $H^{2} = 0$ and $(d/da)H^{2} = 0$ simultaneously ---
separates bounce-permitting from recollapse-only parameter
regimes and provides an analytically tractable handle on the
boundary.

During radiation domination
$f = (8\pi G/3)\rho_{r,0}\,a^{-4}$ and $\Lambda \approx 0$, the
modified Friedmann equation~\eqref{eq:modfried} reduces to
\begin{equation}
H^{2}(a) = \frac{8\pi G\,\rho_{r,0}}{3\,a^{4}}
  - \beta^{2} F(a)^{2}\,.
\label{eq:H2bounce}
\end{equation}
The simultaneous conditions $H^{2} = 0$ and $(d/da)H^{2} = 0$
reduce to
\begin{equation}
-2(1 + x) = n x\,, \qquad
x \equiv (a_{\mathrm{bounce}}/a_{0})^{n}\,.
\label{eq:bouncecond}
\end{equation}
Solving gives $x = -2/(n+2)$, which is positive (so that
$a_{\mathrm{bounce}}$ is real and finite) if and only if
$n < -2$.  The bounce scale factor is
\begin{equation}
a_{\mathrm{bounce}}
  = a_{0}\!\left(\frac{-2}{n+2}\right)^{\!1/n}\!,
\label{eq:abounce}
\end{equation}
whose relation to $a_{0}$ depends on the value of~$n$:
\begin{itemize}
\item $a_{\mathrm{bounce}} < a_{0}$ for $-4 < n < -2$,
\item $a_{\mathrm{bounce}} = a_{0}$ at $n = -4$, and
\item $a_{\mathrm{bounce}} > a_{0}$ for $n < -4$.
\end{itemize} 

The same condition $n < -2$ governs the existence of generic
simple-crossing bounces in the surrounding parameter space.
At small scale factors the NC correction $\beta^{2}F(a)^{2}$
grows as $\beta^{2}(a_{0}/a)^{2|n|}$, while the radiation density
grows as $a^{-4}$; the NC term dominates over radiation as
$a \to 0$ if and only if $2|n| > 4$, i.e.\ $n < -2$.  This is
the condition that the modified Friedmann equation has a
forbidden region at small scale factors, which is necessary for
a bounce.  When $\Lambda_{\mathrm{eff}} > 0$ is included on the
matter side, $H^{2}(a)$ crosses zero once between the forbidden
small-$a$ region and the allowed intermediate-$a$ region; the
crossing has positive slope and $\ddot a > 0$, so re-expansion
is automatic.

The case $n = -4$ is the unique value of $n$ for which the simultaneous conditions bounce
satisfies $a_{\mathrm{bounce}} = a_{0}$ exactly ($x = 1$
in~\eqref{eq:bouncecond}).  At this point the bounce condition
$\beta^{2}F(a_{0})^{2} = (8\pi G/3)\rho_{b}$ with
$F(a_{0}) = 2$ gives a relation among the scales,
\begin{equation}
\rho_{b} = \frac{3\beta^{2}}{2\pi G}\,,
\label{eq:rhob_n4}
\end{equation}
or equivalently
\begin{equation}
a_{0} = \!\left(\frac{2\pi G\,\rho_{b}}{3\beta^{2}}\right)^{\!1/4}\!.
\label{eq:a0nfour}
\end{equation}
If $\rho_{b}$ and $\beta$ are taken at Planck scale,
$\rho_{b} \sim \rho_{P}$ and $\beta \sim t_{P}^{-1}$, then
$a_{0} \sim \ell_{P}$, with the precise numerical coefficient
$(2\pi/3)^{1/4} \approx 1.20$.  The bounce condition therefore
fixes one of the three quantities $\{a_{0}, \beta, \rho_{b}\}$
in terms of the other two; the $n = -4$ model has one fewer
free parameter than its $n \neq -4$ cousins.  The NC correction
grows as $\beta^{2}(a_{0}/a)^{8}$ for $a \ll a_{0}$, faster than
radiation ($\rho \propto a^{-4}$), providing the restoring force
for the bounce.

Re-expansion to a standard cosmological history requires
$\Lambda_{\mathrm{eff}} > 0$, which at Planck-scale~$\beta$
demands the same 122-decimal-place cancellation noted in
Section~\ref{sec:friedmann}.  The $n = -4$ model is therefore
 special --- it reduces the free parameters by one
--- but shares the fine-tuning problem common to all
Planck-scale quantum cosmology proposals.

In the radiation-dominated epoch with $\Lambda \approx 0$, the
right-hand side of the modified Friedmann equation,
$f(a) = (8\pi G/3)\rho_{r,0}\,a^{-4}$, is monotonically
decreasing, while the NC correction $\beta^{2}F(a)^{2}$ is
monotonically increasing for $n > 0$.  The two cross at a scale
factor~$a_{\mathrm{turn}}$ defined by
\begin{equation}
\beta^{2}\!\left[1 + \!\left(\frac{a_{\mathrm{turn}}}{a_{0}}\right)^{\!n}\right]^{\!2}
  = \frac{8\pi G\,\rho_{r,0}}{3\,a_{\mathrm{turn}}^{4}}\,.
\label{eq:aturn}
\end{equation}
For $a > a_{\mathrm{turn}}$, the NC correction exceeds the
matter density and $H^{2} < 0$: the region is classically
forbidden.  The scale~$a_{\mathrm{turn}}$ is therefore a
maximum scale factor for the radiation-dominated phase.  Unlike
the $n<-2$ bounce above, this turning point is
one-sided: the classical solution does not re-expand.  The Big
Bang singularity at $a = 0$ is not resolved classically for
$n > 0$.

Two analytically tractable limits bound~$a_{\mathrm{turn}}$.
When $a_{\mathrm{turn}} \ll a_{0}$ the constant NC term
dominates and $F \approx 1$, giving
\begin{equation}
a_{\mathrm{turn}}^{(0)}
  = \!\left(\frac{8\pi G\,\rho_{r,0}}{3\beta^{2}}\right)^{\!1/4}\!,
\label{eq:aturn0}
\end{equation}
independent of $a_{0}$.  When $a_{\mathrm{turn}} \gg a_{0}$ the
power-law term dominates and $F \approx (a/a_{0})^{n}$, giving
\begin{equation}
a_{\mathrm{turn}}^{(n)}
  = \!\left(\frac{8\pi G\,\rho_{r,0}}{3\beta^{2}}\right)^{\!1/(2n+4)}\!
    a_{0}^{n/(n+2)}\,.
\label{eq:aturnn}
\end{equation}
At $a_{\mathrm{turn}}$ the energy density attains its maximum
value,
\begin{equation}
\rho_{\max} = \frac{3\beta^{2}}{8\pi G}
  \!\left[1 + \!\left(\frac{a_{\mathrm{turn}}}{a_{0}}\right)^{\!n}\right]^{\!2}\!.
\label{eq:rhomax}
\end{equation}
In the constant-NC regime ($a_{\mathrm{turn}} \ll a_{0}$):
$\rho_{\max}^{(0)} = 3\beta^{2}/(8\pi G)$.  For
$\beta = t_{P}^{-1}$ this gives $\rho_{\max} \sim \rho_{P}$ and
maximum temperature
\begin{equation}
T_{\max} \sim \!\left(\frac{3\beta^{2}}{8\pi G}\right)^{\!1/4}\!
  \left(\frac{30\,\hbar^{3}c^{5}}{\pi^{2}g_{*}k_{B}^{4}}\right)^{\!1/4}\!
  \sim T_{P}\,,
\label{eq:Tmax}
\end{equation}
a natural ultraviolet cutoff in the thermal history.

The classical solution therefore has no bounce for $n > 0$:
the universe reaches $a_{\mathrm{turn}}$ and recollapses.  The
constraint function
$C(a) = (9 a^{4}/16\pi^{2}G^{2})[f(a) - \beta^{2}F(a)^{2}]$,
which measures $H^{2}$ in units of the classical momentum
squared, satisfies $C \geq 0$ in the classically accessible
region and $C < 0$ in the forbidden region.  For $n > 0$ with
radiation and~$\Lambda_{\mathrm{eff}}$ in the matter sector, the
function $f(a) - \beta^{2}F(a)^{2}$ is strictly monotonically
decreasing in~$a$: both $f'(a) < 0$ (matter dilutes) and
$-2\beta^{2}FF' < 0$ (NC term grows).  Therefore $C(a)$ has a
unique zero at~$a_{\mathrm{turn}}$, the classically allowed
region is $0 \leq a \leq a_{\mathrm{turn}}$, and the forbidden
region $a > a_{\mathrm{turn}}$ extends to infinity with no
second classical branch on the far side. Whether quantum tunneling can connect this 
forbidden region to a second classical branch is discussed in Section IX.

The deformed Raychaudhuri equation~(\ref{eq:raychaudhuri}) confirms $\ddot{a} <0$ at $a_{\mathrm{turn}}$ 
for $n > 0$, so the scale factor reaches a local maximum and recollapses, in contrast 
to the bounce case where $\ddot{a} > 0$ drives re-expansion.

In summary, the early-universe behavior of the commutation relation is 
divided by the sign of~$n$.  For $n < -2$ the NC correction grows
faster than radiation at small~$a$ and provides a classical
restoring force, producing a non-singular bounce, with $n = -4$ giving rise to a bounce 
exactly at $a_0$.  For $n > 0$ no classical bounce exists: the NC correction
grows with~$a$, $H^{2}(a)$ has a single zero at~$a_{\mathrm{turn}}$,
and the universe reaches a maximum scale factor and recollapses
with no second classical branch available at large~$a$ in the
minimal model.  The intermediate range $-2 \le n \le 0$ is
dynamically uninteresting: the NC correction either is constant
($n = 0$), grows degenerately with radiation ($n = -2$), or grows
too slowly to overtake radiation as $a \to 0$ ($-2 < n < 0$); in
all cases, the Big Bang singularity is not resolved.

\begin{figure*}[!t]
\includegraphics[scale=0.37]{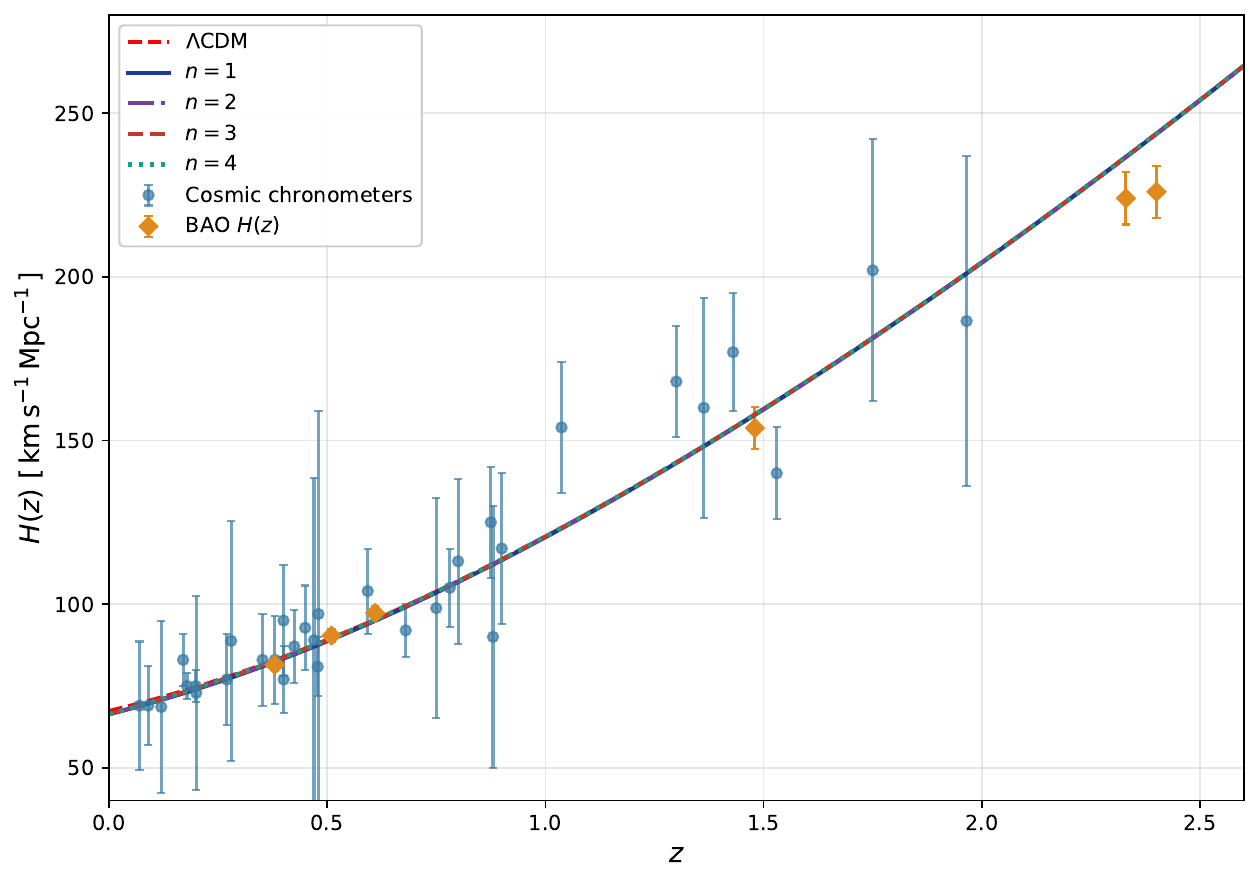}
\includegraphics[scale=0.404]{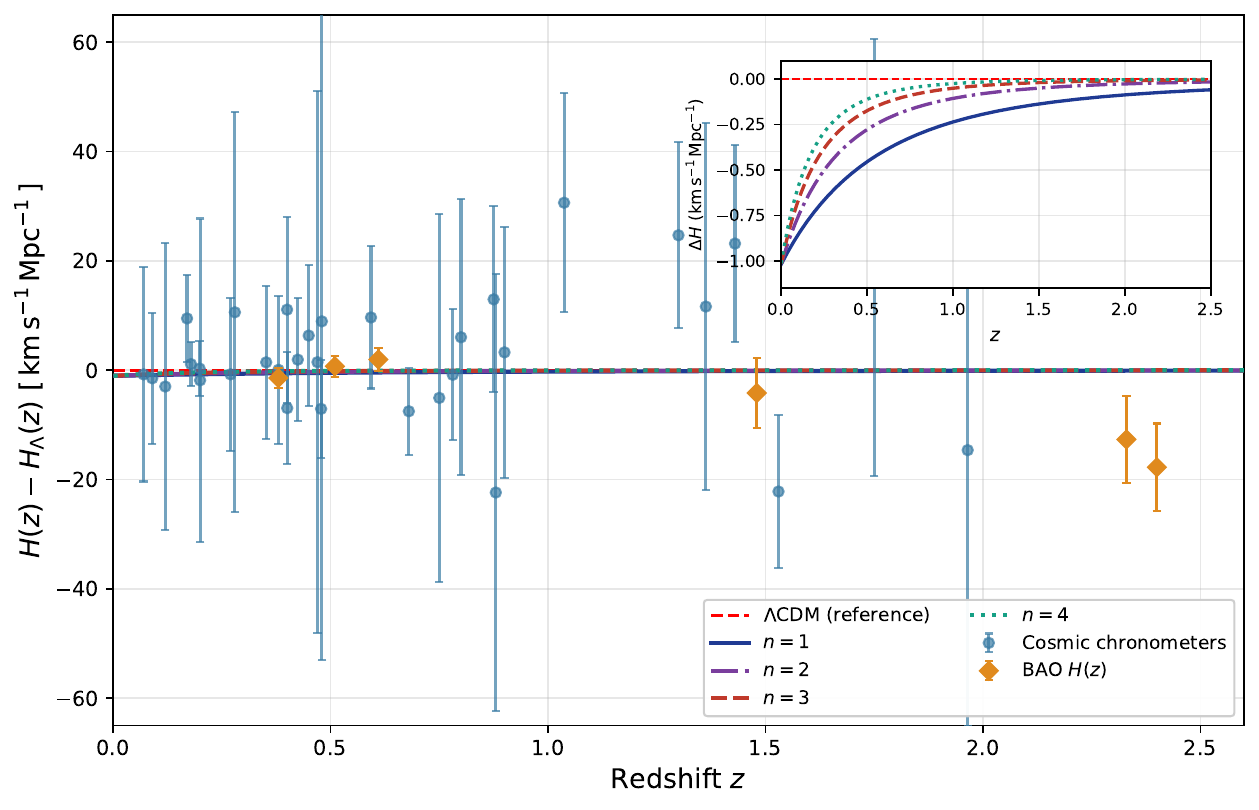}
\caption{{\it Left}: Hubble diagram for the modified Friedmann equation~\eqref{eq:modfried} with
$\varepsilon \equiv \beta/H_0 = 0.1$ and the crossover scale fixed at
$a_0 = a_{\rm today} = 1$. Data points are 32 cosmic chronometer
measurements (filled circles; Moresco et al.~\cite{Moresco2022}) and
6 BAO $H(z)$ measurements (filled diamonds) from BOSS
DR12~\cite{Alam2017}, eBOSS~\cite{Hou2021}, and the BOSS
Lyman-$\alpha$ forest~\cite{Bautista2017,FontRibera2014}. Curves show
the NC model for $n=1$ (navy solid), $n=2$ (purple dash-dotted),
$n=3$ (red dashed), and $n=4$ (teal dotted), together with the
$\Lambda$CDM reference (red dashed). Cosmological parameters
$H_0 = 67.4~{\rm km\,s^{-1}\,Mpc^{-1}}$ and $\Omega_m = 0.315$ are
fixed to Planck 2018 central values~\cite{Planck2018}. 
{\it Right}: Residuals from $\Lambda$CDM, $H(z) - H_\Lambda(z)$, for the
same data and parameters as the left panel.  The inset shows
the same NC curves on a vertical scale matched to their amplitude:
all four converge to $\Delta H \approx -1.0~{\rm km\,s^{-1}\,Mpc^{-1}}$
at $z=0$ (set by the constant NC piece, which is $n$-independent at
$a = a_0$) and decay toward zero at higher redshift, with the decay
rate controlled by $n$. The model deviation is more than an order of
magnitude smaller than typical cosmic-chronometer uncertainties
(median $\sigma_H \approx 17~{\rm km\,s^{-1}\,Mpc^{-1}}$) and remains
below the precision of current BAO measurements; consequently, the
NC model is observationally indistinguishable from $\Lambda$CDM with
present data.
Distinguishing
the NC model from $\Lambda$CDM at the $H(z)$ level will require the
sub-percent precision expected from DESI Year 5 and Euclid.}
\label{fig:hubble}
\end{figure*}

\section{Late-Universe Consequences}
\label{sec:late}

As mentioned earlier (see Section~\ref{sec:earlyuniv}), the model has two 
observationally distinct parameter regimes. The Planck regime was discussed above, 
here we  discuss the  \emph{cosmological regime} ($a_0 \sim H_0^{-1}$, $\beta \ll H_0$) 
where the power-law NC terms
become important near the present epoch.  We focus
on this regime for the observational predictions below.

At late times, radiation and matter dilute and the modified
Friedmann equation reduces to
\begin{equation}
H^{2} = \frac{\Lambda_{\mathrm{eff}} c^{2}}{3}
  - 2\beta^{2}\!\left(\frac{a}{a_0}\right)^{\!n}
  - \beta^{2}\!\left(\frac{a}{a_0}\right)^{\!2n}\!.
\label{eq:Hlate}
\end{equation}
The two NC power-law terms grow for $n > 0$, progressively
reducing $H^{2}$ and acting as dark energy with $w>-1$,
while for $n \le 0$ they dilute and
$H^{2} \to \Lambda_{\mathrm{eff}} c^{2}/3$, approaching de~Sitter
with a renormalized cosmological constant.

Setting $a_0 = a_{\mathrm{today}} = 1$ and $\beta = \epsilon H_0$, 
the dimensionless Hubble rate  (see
Fig.~\ref{fig:hubble}) is 
\begin{align}
E^{2}(z) &\equiv \frac{H^{2}(z)}{H_0^{2}}
  = \Omega_r(1+z)^{4} + \Omega_m(1+z)^{3} + \Omega_\Lambda
  \notag\\
  &\quad - \epsilon^{2}\!\left[2(1+z)^{-n} + (1+z)^{-2n}\right],
\label{eq:Hz_general}
\end{align}
where the constant NC term has been absorbed into
$\Omega_\Lambda$.  The NC correction is largest today for $n > 0$
and grows with redshift for $n < 0$.  The fractional deviation
from $\Lambda$CDM is
\begin{equation}
\frac{E^{2}(z) - E^{2}_{\Lambda\mathrm{CDM}}(z)}{H_0^{2}}
  = -\epsilon^{2}\!\left[2(1+z)^{-n} + (1+z)^{-2n}\right].
\label{eq:Hz_deviation}
\end{equation}
For $n < 0$ the dominant
growing piece is $(1+z)^{2|n|}$ (from the $-2n$ power), while
for $n > 0$ both terms decay with redshift.  The power-law form
is structurally distinct from the $w_0 w_a$ Chevallier–Polarski–Linder (CPL) parametrization,
in which the deviation is linear in~$a$; the shape of $H(z)$ is
therefore a discriminant between the two.

\begin{figure}[t]
\includegraphics[width=\linewidth]{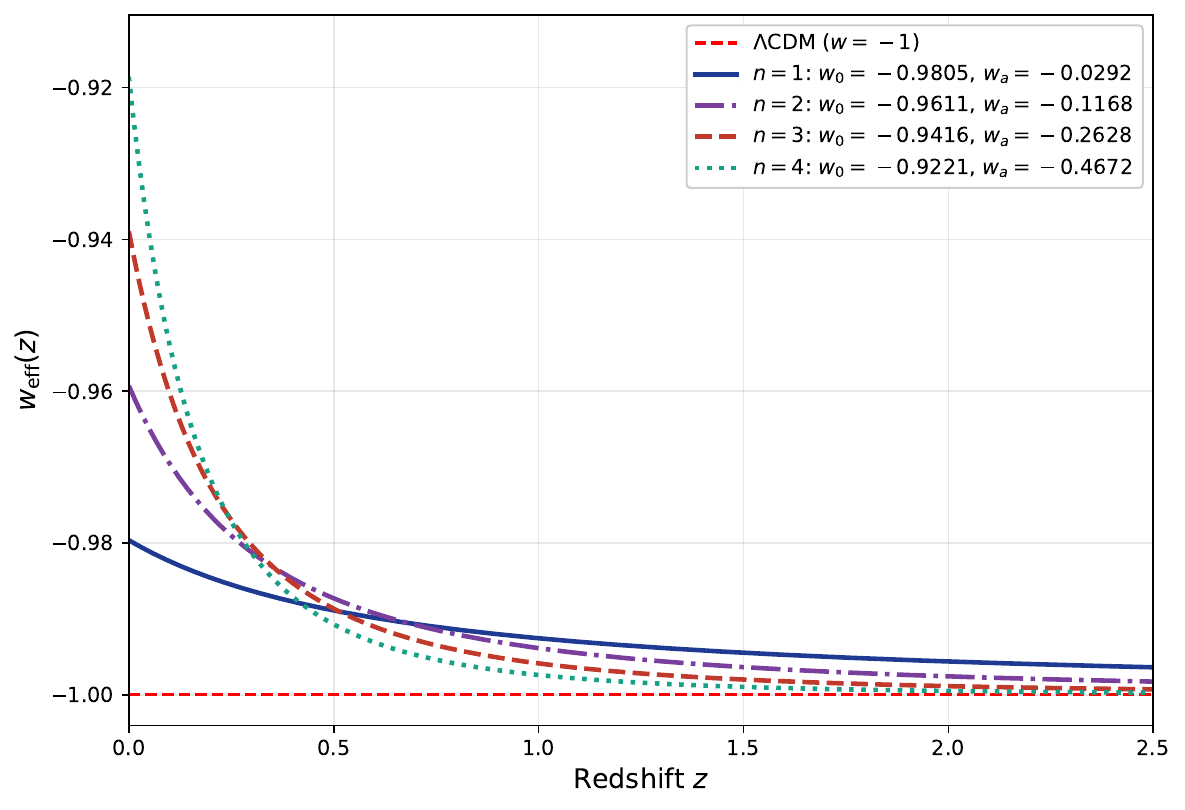}
\caption{Effective dark energy equation of state $w_{\rm eff}(z)$
from Eq.~(\ref{eq:weff}) at $\varepsilon = 0.1$ and
$a_0 = a_{\rm today} = 1$, for $n=1$ (navy solid), $n=2$ (purple
dash-dotted), $n=3$ (red dashed), and $n=4$ (teal dotted). The
dashed red line marks $w=-1$. CPL coefficients $(w_0, w_a)$ from
Eqs.~(\ref{eq:w0})--(\ref{eq:wa}) are reported in the legend.
For all $n>0$, $w_{\rm eff} > -1$ at every redshift --- a structural
prediction of the commutation relation, since the numerator
$2n\varepsilon^2 a^n(1+a^n)$ in Eq.~(\ref{eq:weff}) is
positive-definite for $n>0$. The curves peak at $z=0$, where the NC
contribution is largest, and asymptote to $w=-1$ at high redshift as
the power-law term $(a/a_0)^n \to 0$. This redshift dependence is
structurally distinct from CPL, in which $w(z) \to w_0 + w_a$ at
high $z$; the model becomes increasingly $\Lambda$-like at early
times rather than asymptoting to a fixed phantom or quintessence
value. The crossing of the $n$-curves near $z \approx 0.5$ is a
further signature absent from any single-parameter CPL fit. }
\label{fig:weff}
\end{figure}

The effective dark energy equation of state
(Fig.~\ref{fig:weff}) is
\begin{equation}
w_{\mathrm{eff}}(a)
  = -1 + \frac{2n\epsilon^{2} a^{n}(1 + a^{n})}
    {3\,\Omega_{\mathrm{DE}}(a)}\,,
\label{eq:weff}
\end{equation}
where $\Omega_{\mathrm{DE}}(a) = \Omega_\Lambda
- \epsilon^{2}(2a^{n} + a^{2n})$.  The CPL approximation at
$a = 1$ gives
\begin{align}
w_0 &\approx -1 + \frac{4n\epsilon^{2}}{3\Omega_\Lambda}\,,
\label{eq:w0}
\\
w_a &\equiv -\frac{\partial w}{\partial a}\bigg|_{a=1}
  \approx -\frac{2n^{2}\epsilon^{2}}{\Omega_\Lambda}\,.
\label{eq:wa}
\end{align}
Since $n^{2} > 0$ always, $w_a < 0$ for all $n \neq 0$; the sign
of $w_0 - (-1)$ directly reveals the sign of~$n$.  DESI Year~1
BAO data prefer $w_0 \approx -0.55$ to $-0.73$ and
$w_a \approx -0.7$ to $-1.3$~\cite{DESI2024}, though the robustness of this 
dynamical dark energy signal across DR1 and DR2 has been questioned~\cite{Colgain:2025nzf}.  
The $n>0$ cosmological regime predicts $w_0 > -1$ and $w_a < 0$, matching the signs 
of both DESI preferences, though with percent magnitudes for $\epsilon \lesssim  0.1$. 
The decisive test will come from the shape of $H(z)$
(Eq.~\eqref{eq:Hz_deviation}) where the NC model could potentially be distinguished from the
CPL form by DESI~Year~5 and Euclid.

Current BAO and supernova data constrain the comoving distance
shift.  At lowest order in $\epsilon^{2}$ and for
$|n| \gtrsim 0.5$, the fractional shift scales as
\begin{equation}
\frac{\delta d_C(z)}{d_C^{\Lambda\mathrm{CDM}}(z)}
  \;\sim\; \frac{\epsilon^{2}}{n(1+n)}\,,
\label{eq:dC_shift}
\end{equation}
with an $\mathcal{O}(1)$ prefactor that depends on the redshift
range integrated.  Percent-level consistency with present data
requires $\epsilon^{2} \lesssim \mathcal{O}(10^{-2})$ for
$a_0 \sim H_0^{-1}$.  For $n > 0$, expansion halts at
\begin{equation}
a_{\max} \approx a_0\,\epsilon^{-1/n}\,,
\label{eq:amax_cosmo}
\end{equation}
with $a_{\max}/a_0 = 10, 3.2, 2.2$ for $\epsilon = 0.1$ and
$n = 1, 2, 3$, corresponding to future collapse timescales of
roughly $30$, $10$, and $5$~Gyr.  Models with
$\epsilon \gtrsim 0.1$ and $a_0 \lesssim a_{\mathrm{today}}$ are
already excluded by the observed expansion to $z \sim 2.3$.

For $n < 0$, the universe expands indefinitely toward de~Sitter,
but BBN restricts the viable range of $|n|$.  The NC
correction scales as $\epsilon^{2}(1+z)^{2|n|}$ and must not
disrupt expansion at BBN~\cite{KolbTurner1990,Steigman:2007xt} ($z \sim 10^{10}$).  Requiring the NC
correction to remain subdominant to radiation at BBN gives
$\epsilon^{2}(1+z_{\mathrm{BBN}})^{2|n|}
\lesssim \Omega_r(1+z_{\mathrm{BBN}})^{4}$, which yields
\begin{equation}
\epsilon^{2} \;\lesssim\; \Omega_r\,(1+z_{\mathrm{BBN}})^{4-2|n|}\,.
\label{eq:BBN}
\end{equation}
For $|n| \lesssim 0.1$ the constraint is harmless and the NC
correction is observable at low redshift.  For
$|n| \gtrsim 2$ the constraint tightens rapidly: at $|n| = 4$
it forces $\epsilon \lesssim 10^{-22}$, making the present-epoch
correction unobservable.  Only $|n| \lesssim 0.1$ is both consistent with 
BBN and observationally accessible.

Note that although the {\it individual} power-law components have formal equations of state 
$w_1, w_2 < -1$ for $n > 0$ (see~\eqref{eq:F2expand}) -- even though the {\it combined} 
effective dark energy satisfies $w_{\mathrm{eff}} > -1$ -- the model does not propagate a 
ghost. In scalar-field phantom models, instability
arises from a wrong-sign kinetic term for a new degree of
freedom; here no such field exists.  The NC correction is a
geometric term on the left-hand side of the Friedmann equation,
sourced by the deformed phase-space structure; the matter action,
kinetic terms, and equations of motion are identical to standard
cosmology.  At leading order in $\beta/H$ the NC deformation does
not enter the matter perturbation equations, so the sound speed
of scalar perturbations is unchanged.  These are qualitative
arguments; a rigorous analysis of the perturbed action is left
for future work.

The NC suppression factor,
\begin{equation}
\Xi_n(a) \equiv \frac{\beta^{2}(1 + (a/a_0)^{n})^{2}}
  {(8\pi G/3)\rho + \Lambda c^{2}/3}
  = 1 - \frac{H^{2}}{H^{2}_{\mathrm{tot}}}\,,
\label{eq:Xi}
\end{equation}
measures the fractional suppression of $H^{2}$ by the NC
correction.  At horizon crossing in the cosmological regime,
$\Xi_n(a_*) \approx \beta^{2}/H^{2}(a_*) \ll 1$, so the NC
correction is negligible on all scales accessible to CMB
experiments.  \emph{The model therefore is consistent with  a standard
inflationary power spectrum}; all observable NC signatures are
confined to~$H(z)$.  Near the classical turning points
$\Xi_n \to 1$ and the perturbative expansion breaks down; a
non-perturbative treatment is left for future work.

\section{Relation to Prior Work}
\label{sec:related}

The prior noncommutative cosmology literature falls into several
distinct classes, each deforming a different piece of the
minisuperspace structure: the canonical
configuration--momentum bracket, the bracket between the scale
factor and an auxiliary scalar field, the bracket between
momenta, or dispersion relations in the matter sector.  The
present proposal sits in none of these classes.  It deforms the
velocity--configuration bracket $[\hat a, \hat{\dot a}]$
directly, a structure that to our knowledge has not been studied 
in prior work.

A series of recent works have applied GUP-modified cosmology to current observational data, 
deriving modified Friedmann or Raychaudhuri equations and confronting them with DESI BAO, 
Pantheon+ supernovae, and redshift-space distortion 
measurements~\cite{Paliathanasis:2025mfj,Paliathanasis:2025dcr,Paliathanasis:2025kmg,Paliathanasis:2026vhi}. 
These papers deform the canonical $[\hat{a}, \hat{p}_a]$  bracket, structurally distinct from the velocity-
bracket deformation of the present work as discussed in Section~\ref{sec:algebra_rep}.

Vakili et al.~\cite{Vakili2008} and subsequent
work~\cite{Battisti2008,Bosso2020} deform the algebra
$[\hat a,\hat p_a]=i\hbar(1+\alpha\hat p_a^{2}/M_P^{2}c^{2})$,
yielding a modified Friedmann equation with the NC correction on
the right-hand side as a multiplicative factor in the energy
density, $\sim 1 + \alpha\rho/\rho_P$.  
This is structurally
distinct from what we propose here: our correction
is on the left-hand side and
grows as a power of~$a$ rather than of~$\rho$.

Garc\'ia-Compe\'an et al.~\cite{GarciaCompean2002} and Guzman et
al.~\cite{Guzman2007} introduced $[\hat a,\hat\phi]=i\theta$ for
a scalar field $\phi$ coupled to gravity.  Their noncommutativity
is between two different degrees of freedom ($a$ and $\phi$),
whereas what we propose here deforms the algebra of~$a$ with itself (its
velocity), and no scalar field is required.  Their NC correction
is linear in $\theta$ and perturbative; ours is
non-perturbative and exact.

Models deforming
$[\hat p_a,\hat p_\phi]=i\eta$~\cite{Obregon2001,Vakili2010,
DiazBarron2021} act on momenta rather than on configuration-space
variables ($a$ and $\dot a$), and require a scalar field to
produce non-trivial dynamics.  A related precedent is the work
of Freidel and Livine~\cite{FreidelLivine2006}, who showed that
integrating out gravitational degrees of freedom in 3D quantum
gravity yields an effective noncommutative quantum field theory
with $\kappa$-deformed Poincar\'e symmetry.  That result
demonstrates a direct path from a gravitational theory to
noncommutative phase-space structure, providing a concrete
microscopic mechanism of the kind that motivates the present
proposal, though in a different spacetime dimension and with a
different algebra.

The LQC bounce arises from holonomy corrections in the
Hamiltonian~\cite{Ashtekar2006,Singh2009}, giving
$H^{2} = (8\pi G/3)\rho(1 - \rho/\rho_c)$~\cite{Ashtekar2011}.  The LQC
correction has a fixed functional form (quadratic in $\rho$) and
appears on the right-hand side.  In the present model the turning
point occurs when $\beta^{2}(1+(a/a_0)^{n})^{2}$ equals the energy
density, a geometric condition on~$a$ rather than~$\rho$.
For $n = -4$ the present model produces a genuine classical
bounce at Planck density through an NC-driven acceleration
mechanism; the LQC and $n = -4$ bounces are physically distinct
in their origin but both produce a classical re-expansion at
Planck density.  The proposed model studied in this paper  requires an external
fine-tuning of~$\Lambda$ for the Planck-regime bounce to connect
to a standard cosmological history, while LQC avoids this because
the critical density dilutes on the right-hand side.  For a
comprehensive review of bouncing cosmologies as alternatives to
inflation, and the challenges facing each implementation, see
Brandenberger and Peter~\cite{BrandenbergerPeter2017}.

Rasouli et al.~\cite{Rasouli2011,Rasouli2014} introduced
noncommutativity between the scale factor and the Brans--Dicke
scalar field.  This is the closest structural analogue to the
present proposal, in that the deformation acts on a bracket
involving~$a$.  The difference is that their deformation involves
a second field, whereas ours involves only~$a$ and its velocity, without
additional matter fields.

Modifying dispersion relations through $\kappa$-Poincar\'e
algebra deformations~\cite{AmelinoCamelia2001,MagueijoSmolin2002}
leads to modified equations of state for radiation embedded in
the LQC or standard Friedmann framework~\cite{Gubitosi2023}.  The
resulting modifications to the Hubble rate at early times are
qualitatively similar to the $n < 0$ regime of the present model.
The physical origin is different: in the $\kappa$ approach the
modification comes from the matter sector, whereas here it comes
purely from the cosmological uncertainty principle. 

Alexander, Brandenberger, and
Magueijo~\cite{AlexanderBrandenbergerMagueijo2003} showed that a
radiation-dominated universe subject to spacetime quantization
--- via deformed dispersion relations with a maximal momentum ---
develops a trans-Planckian branch in which radiation acquires
negative pressure, driving an inflationary equation of state
without an inflaton field.  This model, developed alongside the
companion paper by Alexander and
Magueijo~\cite{AlexanderMagueijo2001} on NC geometry as a varying
speed of light, connects NC spacetime structure directly to
early-universe dynamics.  The present model shares the motivation
(NC geometry as the source of modified early-universe expansion)
but differs fundamentally in mechanism: the deformation acts on $[\hat a, \hat{\dot a}]$
rather than on the matter dispersion relation, and the resulting
NC correction is geometric (left-hand side of the Friedmann
equation) rather than thermodynamic (equation of state of
radiation).

\section{Discussion}
\label{sec:discussion}

In this section we discuss the limitations, open questions, and the interpretation of the proposed cosmological commutation relation.

\subsection*{Limitations}
\label{sec:discussion_limits}

The central limitation of the model is the constant NC
term~$\beta^{2}$, which shifts the effective cosmological constant by
$\Lambda_{\mathrm{eff}} = \Lambda - 3\beta^{2}/c^{2}$. For $\beta
\sim t_{P}^{-1}$ this requires a 122-decimal-place cancellation;
the model does not resolve the cosmological constant problem. Three
paths exist within the present framework: (i)~accept the fine-tuning,
as in any Planck-scale model (the $n = -4$ bounce fits this category);
(ii)~work in the cosmological regime $\beta \ll H_{0}$, where no
fine-tuning is needed and the model is a predictive two-parameter
dark energy framework; or (iii)~modify the commutation relation to
remove the constant term, at the cost of losing the two-regime
crossover structure.

For $n > 0$ a structural no-go result further restricts the model:
no single choice of $a_{0}$ and $\beta$ simultaneously produces a
Planck-scale early-universe turning point, an observable
late-universe dark energy signal, and BBN-consistent expansion through
the present epoch. With $a_{0} = \ell_{P}$ and $\beta = t_{P}^{-1}$,
Eq.~(\ref{eq:amax_cosmo}) gives a late-universe turning point at
$a_{\max} \ll \ell_{P}$, unphysical. With $a_{0} \sim H_{0}^{-1}$ and
$\beta \sim H_{0}$, the constant NC term $\beta^{2} \sim H_{0}^{2}$
halts radiation-dominated expansion at $z \sim 20$, so consistency with BBN 
forces $\beta^{2}/H_{0}^{2} \ll \Omega_{r} \sim 10^{-5}$, suppressing
the present-epoch power-law correction to $\lesssim 10^{-5}$, which is
undetectable with any foreseeable survey. The $n > 0$ family is
therefore best understood as either a late-universe dark energy
model (cosmological regime) or an early-universe quantum cosmology
model (Planck regime), but not both.

Moreover, for $n > 0$ in the minimal radiation plus
$\Lambda_{\mathrm{eff}}$ matter content, the constraint function
$f(a) - \beta^{2}F(a)^{2}$ is monotonically decreasing in~$a$, so
$C(a)$ has a single classical zero at~$a_{\mathrm{turn}}$ and the
forbidden region extends to large~$a$ without a second classical
branch on the far side. The Big Bang singularity is therefore not
resolved classically for $n > 0$, and no Hartle--Hawking or
Vilenkin tunnelling can give rise to the universe within this minimal
model: there is no allowed region at large~$a$ for the wave function
to tunnel into. Singularity resolution for $n > 0$ would require
either matter content beyond radiation and $\Lambda_{\mathrm{eff}}$
or a modification of the commutation relation such that $f -
\beta^{2}F^{2}$ becomes non-monotonic.

For $n < -2$ the commutation relation supports a non-singular bounce
at small scale factors, with $n = -4$ giving rise to a bounce
condition that fixes $a_{0}$ in terms of $\beta$ and the bounce
density. The Planck-scale fine-tuning of the cosmological constant
persists in this regime.

A further limitation concerns the $H_{0}$ tension (see
e.g.,~\cite{Kamionkowski:2022pkx}). The NC correction enters the
Friedmann equation as $+\beta^{2}F^{2}$ on the left-hand side,
reducing the expansion rate available from a given energy budget.
For Planck-anchored $\Omega_{m}$ and $r_{s}$, this lowers the
inferred $H_{0}$ relative to $\Lambda$CDM by roughly $1.5\%$ at
$\epsilon = 0.1$ --- a shift in the opposite direction with respect
to the late-universe SH0ES value. This is a generic result: $F^{2}$
is positive-definite, so the NC term subtracts from $H^{2}$
regardless of the sign of $n$, and for $n > 0$ the effective dark
energy is quintessence-like ($w_{\mathrm{eff}} > -1$), which is known
to worsen rather than alleviate the tension~\cite{Lee:2022cyh}. The model therefore
does not address the $H_{0}$ tension, and a resolution within this
framework would require either modifying the sign structure of the
deformation or coupling it to early-universe physics that reduces
the sound horizon --- neither of which is present in this minimal
model.

\subsection*{Open questions}

Several directions remain open:
(i)~whether extending the matter content or modifying the commutation
relation such that $f - \beta^{2}F^{2}$ becomes non-monotonic can
allow a tunnelling channel of the
Hartle--Hawking~\cite{HartleHawking1983} or
Vilenkin~\cite{Vilenkin1984} type, and what selection effects such a
channel would impose;
(ii)~the perturbative stability of the classical bounce for $n < -2$
under metric and matter back-reaction is unestablished;
(iii)~the equations become non-perturbative near the bounce where
$H \to 0$ and $\Xi_{n} \to 1$; and
(iv)~a systematic Bayesian analysis confronting
Eq.~\eqref{eq:Hz_general} with data from DESI, Euclid, and LSST would
constrain $(\epsilon, n)$.

The proposed commutation relation~\eqref{eq:algebra} is not
covariant under the full group of spacetime diffeomorphisms: the
factor $\hat a^{2}$ transforms non-trivially under a change of time
slicing, making comoving gauge the appropriate frame for its
formulation (similar to LQC~\cite{Cailleteau2012}). A fully
gauge-invariant formulation of the deformed algebra remains an
important open problem.  
However, a  promising direction is to recast the deformation in constant-mean-curvature 
(York-time) variables~\cite{York1972}, where the relation becomes a deformation of the bracket between 
the spatial volume and the York time---both gauge-invariant quantities---suggesting that perhaps the 
apparent non-covariance is an artifact of the comoving-gauge variables rather than an intrinsic 
feature of the deformation~\cite{GomesGrybKoslowski2011}. 

The most fundamental open question is the microscopic origin of the
algebra. Several research programs provide candidate ingredients.
Group Field Theory~\cite{Freidel2005gft, Oriti2009, Gielen2014gft, Oriti2007}
produces an emergent scale-factor commutator with an $a^{2}$ leading
dependence consistent with the $\hat a^{2}$ structure of the present
algebra, with $n$ running under the GFT renormalisation
group~\cite{Gielen2013, Oriti2016, Carrozza2016}. Other programs
--- polymer and loop quantisation~\cite{BenAchour2019polymer,
Corichi2007,Veneziano1991}, string T-duality~\cite{GasperiniVeneziano2003}, and
modified horizon thermodynamics~\cite{Jacobson1995, Barrow2020,Nojiri2022} ---
contain noncommutative phase-space structures with qualitative
resemblances to Eq.~\eqref{eq:algebra}. None of these frameworks,
however, derives Eq.~\eqref{eq:algebra} exactly.

A unifying picture emerges if one promotes $n$ to $n(a)$ as a running
coupling. The deformation then passes through the bounce phase
($n < -2$), a transition ($-2 < n < 0$), a $\Lambda$-like phase
($n \approx 0$), and the dark energy phase ($n > 0$). Whether any
formal theory actually produces such a trajectory --- interpolating
from $(n \approx -4, a_{0} \sim \ell_{P}, \beta \sim t_{P}^{-1})$ at
the bounce to small positive~$n$ at $a_{0} \sim H_{0}^{-1}$ today ---
is an open question.

\subsection*{Quantum gravity at cosmological scales}
\label{sec:UV_IR}

The apparent tension between the quantum-gravitational motivation of
the modified Friedmann equation~\eqref{eq:modfried} and its
observationally viable regime rests on an assumption that is worth
questioning. The standard expectation is that quantum-gravitational
effects are confined to the very early universe ($a \to 0$),
dominating near the Big Bang and decoupling at late times. This
assumption is grounded in dimensional analysis: quantum-gravitational
corrections are presumed to be suppressed by powers of $\ell_{P}/
\lambda$ for any physical length~$\lambda$. The expectation that
quantum gravity decouples at low energies works for ordinary quantum
field theory, where corrections shrink as one moves away from the UV
cutoff. The kinematics of the whole universe is a different problem.
The commutation relation~\eqref{eq:algebra} applies at any value of
$a$, and the scale at which it matters depends on $\beta$ and
$a_{0}$, values that do not have to be at the Planck scale.

A more coherent view emerges from the holographic principle. UV/IR
mixing is a generic feature of noncommutative field
theories~\cite{Seiberg1999,Connes1994} and is expected in any consistent
quantum theory of gravity through the holographic
bound~\cite{Jacobson1995,CaiKim2005}. Freidel et
al.~\cite{Freidel2021infrared,Freidel2022UVIR,AmelinoCamelia2011rl}
have argued that quantum-gravity effects may leave observable
signatures in the infrared. From this perspective, $[\hat a,
\hat{\dot a}] \neq 0$ is a statement about the cosmological horizon
rather than about the Planck scale. It encodes a quantum property
of the horizon, with~$\beta$ set by the horizon scale in direct
analogy with the Hawking temperature: for a black hole, $T_{H}$ is
set by the surface gravity of the horizon, not by the Planck
temperature; a Planck-sized black hole has $T_{H} \sim T_{P}$,
while a solar-mass one has $T_{H} \sim 10^{-8}$~K. Applying the same
logic here, $\beta \sim c/a_{\mathrm{horizon}}$: when the
cosmological horizon is Planck-sized, $\beta \sim t_{P}^{-1}$; when
the horizon is of order $H_{0}^{-1}$, $\beta \sim H_{0}$. Modified
horizon thermodynamics~\cite{Barrow2020,Tsallis2013} produces
Friedmann-equation corrections of qualitatively similar form.

This reverses the standard hierarchy of scales. What is
conventionally treated as fundamental --- the Planck scale --- is
now a special case, and what is conventionally treated as derived
--- the horizon --- is fundamental. The cosmological regime $a_{0}
\sim H_{0}^{-1}$, $\beta \sim H_{0}$ is not a low-energy limit of a
fundamentally Planck-scale theory; it is the generic application of
a commutation relation whose fundamental parameters track the
horizon. On this view, the late-universe phenomenology is not a
marginal signature of deeper physics decoupled at the Planck scale;
it is the primary observational consequence of the commutator
operating at that scale.

The holographic interpretation also suggests a specific restriction on the
parameter space. If $\beta$ tracks the horizon scale, the natural
relation is $\beta = c/a_{0}$, in direct analogy with the
Hawking temperature $T_{H} \sim c/r_{S}$ for a black hole. This
collapses the three-parameter family $(\beta, a_{0}, n)$ to a
two-parameter family $(a_{0}, n)$: the deformation strength is
determined by the crossover scale via the speed of light, and only
the crossover scale and the exponent remain free.

Under this restriction, the constant NC contribution to the
cosmological constant becomes $\Lambda_{\mathrm{eff}} = \Lambda -
3/a_{0}^{2}$. In the cosmological regime $a_{0} \sim c/H_{0}$, this
is of order $H_{0}^{2}$, comparable in magnitude to the observed
$\Lambda$ --- the 122-decimal-place fine-tuning of the unconstrained
theory is replaced by a natural-scale shift. This does not solve 
the cosmological constant problem. Instead of explaining why
$\Lambda$ is small, the model now requires explaining why $a_{0}$
tracks the cosmological horizon (the holographic argument above 
suggests one natural answer). The
Planck-regime fine-tuning persists for $a_{0} = \ell_{P}$, since the
constraint then forces $\beta = c/\ell_{P} = t_{P}^{-1}$
automatically, and does not address the hierarchy between $H_{0}$
and $t_{P}^{-1}$.

\section{Conclusions}
\label{sec:conclusions}

In summary, we proposed a non-zero cosmological commutation relation
between the scale factor and its rate of expansion,
$[\hat a, \hat{\dot a}] = -i\beta\hat a^{2}[1 + (\hat a/a_{0})^{n}]$,
and derived its cosmological consequences across the full range
of~$n$. The modified Friedmann equation~\eqref{eq:modfried} produces
three additional terms: a constant shift that renormalises $\Lambda$,
and two power-law terms with equations of state $w_{1} = -1 - n/3$
and $w_{2} = -1 - 2n/3$ whose behavior is determined by the sign
and magnitude of~$n$. For $n < -2$ the same deformation produces a
classical bounce at $a_{\mathrm{bounce}} = a_{0}[-2/(n+2)]^{1/n}$,
with $n = -4$ giving rise to a bounce that coincides with the NC
crossover scale $a_{0}$. For $n > 0$ the NC correction grows
with~$a$, producing a classical maximum scale factor
$a_{\mathrm{turn}}$ and an effective dark energy with
$w_{\mathrm{eff}} > -1$ at late times, with all observable signatures
confined to measurements of~$H(z)$.

Quantum gravity is conventionally expected to appear at the Planck
scale and decouple from observable physics elsewhere. The model
studied here suggests an alternative: a quantum kinematic structure
whose deformation parameter is set not by a fixed fundamental length
but by the horizon --- the Planck regime is the special case of an
early universe whose horizon was Planckian, while the cosmological
regime corresponds to the horizon size today. If $H(z)$ deviates
from $\Lambda$CDM in the power-law form predicted by
Eq.~\eqref{eq:Hz_general}, it may be a measurement of the
deformation parameter at the cosmological horizon. In this picture,
the accelerated expansion of the late universe may be the
macroscopic imprint of an irreducible quantum uncertainty in
simultaneously knowing the size and expansion rate of the universe.

\begin{acknowledgments}

We thank the anonymous referee for constructive feedback. We acknowledge 
useful conversations with Stephon Alexander, Robert Brandenberger, 
Laurent Freidel, Cristiano Germani, Antal Jevicki, David Kagan, Rocky Kolb, 
Jo\~ao Magueijo, and Andrew Zentner. We thank the organizers of the Cosmo2025, 
the Aspen Center for Physics, and the 2026 Caribbean Future of Science Symposium 
where part of this work was completed. We acknowledge the use of the Claude AI 
Assistant (Anthropic) for assistance in checking the calculations in this manuscript. 
The author takes sole responsibility for all content. This work was supported by 
National Science Foundation (NSF) No. PHY-2412666.
\end{acknowledgments}

\end{document}